\documentclass[aps,prb,twocolumn,superscriptaddress,floatfix]{revtex4}

\usepackage{amsmath,amssymb,bm}
\usepackage{epsfig}
\usepackage{graphics}

\newcommand{\bk}{{\bf k }}

\newcommand{\bq}{{\bf q }}
\newcommand{\br}{{\bf r }}

\newcommand{\EP}{{\it e}-ph}

\begin{document}

\title{Electron-Phonon Interactions in Graphene, Bilayer Graphene, and Graphite}

\author{Cheol-Hwan Park}
\email{cheolwhan@civet.berkeley.edu}
\affiliation
{Department of Physics, 
University of California at Berkeley, 
Berkeley, California 94720, USA}
\affiliation
{Materials Sciences Division, 
Lawrence Berkeley National Laboratory, Berkeley, 
California 94720, USA}
\author{Feliciano Giustino$^\dagger$}
\affiliation
{Department of Physics, 
University of California at Berkeley, 
Berkeley, California 94720, USA}
\affiliation
{Materials Sciences Division, 
Lawrence Berkeley National Laboratory, Berkeley, 
California 94720, USA}
\author{Marvin L. Cohen} 
\affiliation
{Department of Physics, 
University of California at Berkeley, 
Berkeley, California 94720, USA}
\affiliation
{Materials Sciences Division, 
Lawrence Berkeley National Laboratory, Berkeley, 
California 94720, USA}
\author{Steven G. Louie}
\affiliation
{Department of Physics, 
University of California at Berkeley, 
Berkeley, California 94720, USA}
\affiliation
{Materials Sciences Division, 
Lawrence Berkeley National Laboratory, Berkeley, 
California 94720, USA}

\begin{abstract}
Using first-principles techniques, we calculate
the renormalization of the electron Fermi velocity and
the vibrational lifetimes arising from electron-phonon
interactions in doped bilayer graphene and in graphite
and compare the results with the corresponding quantities in graphene.
For similar levels of doping, the Fermi velocity renormalization
in bilayer graphene and in graphite
is found to be approximately 30\% larger than that in graphene.
In the case of bilayer graphene, this difference
is shown to arise from the interlayer interaction.
We discuss our findings in the light of recent
photoemission and Raman spectroscopy experiments.
\end{abstract}

\date{\today}

\maketitle

Since the fabrication of crystalline graphitic films
with a thickness of only
a few atoms~\cite{novoselov:2004Sci_Graphene,
novoselov:2005PNAS_2D,novoselov:2005Nat_Graphene_QHE,
zhang:2005Nat_Graphene_QHE,berger:2006Sci_Graphene_Epitaxial},
single- and double-layer graphene have received
considerable attention~\cite{geim:2007NatMat_Graphene_Review}.
These materials are promising
candidates for nanoelectronics
applications because of the high mobility of charge
carriers in these systems and the tunability of their electronic properties
by gating~\cite{geim:2007NatMat_Graphene_Review}.
Since electron-phonon (\EP) interaction plays an important
role in the dynamics of charge
carriers~\cite{engelsberg:1963PR_ElPh,grimvall:1981_Metal_ElPh},
understanding its effects in single- and double-layer graphene
is of crucial importance for graphene-based electronics.

The \EP\ interaction in metals
modifies the dynamics of electrons with energy near the Fermi level
by increasing their mass and reducing their lifetime.
The mass renormalization can be described in
terms of the \EP\ coupling strength $\lambda_{n\bk}$,
defined as the energy derivative of the real part of the
phonon-induced electronic self-energy $\Sigma_{n\bk}(E)$ at the Fermi level $E_{\rm F}$:
$\lambda_{n\bk}=-\partial\,{\rm Re}\Sigma_{n\bk}(E)/\partial E|_{E=E_{\rm F}}$,
where $n$, $\bk$ and $E$ are the band index, the wavevector and 
the energy of the electron, respectively~\cite{grimvall:1981_Metal_ElPh}.
The electron mass renormalization can be obtained from the \EP\
coupling strength through $m^*/m=1+\lambda_{n\bk}$ where $m$ and $m^*$ are
the bare band mass and the renormalized mass, respectively.
The \EP\ interaction also gives rise to a
phonon lifetime $\tau_{\nu\bq}=\hbar/\Gamma_{\nu\bq}$~\cite{PhysRevLett.29.1593},
where $\Gamma_{\nu\bq}$ is the phonon linewidth,
i.\,e.\,, twice
the imaginary part of the phonon self-energy
arising from the \EP\ interaction.
Here $\nu$, $\bq$ and $\omega$ are
the phonon branch index, the wavevector and the
energy of the phonon, respectively~\cite{grimvall:1981_Metal_ElPh}.

These quantities can be calculated from first-principles
within the Migdal approximation as~\cite{grimvall:1981_Metal_ElPh}
\begin{eqnarray}
&&\lambda_{n\bk} 
=\sum_{m,\nu} \, \int \, \frac{d\bq}{A_{\rm BZ}}
\; |g_{mn,\nu}(\bk,\bq)|^2 \nonumber \\
&&\times\left[\frac{n_{\bq\nu}
+1-f_{m\bk+\bq}}
{(E_{\rm F}-\epsilon_{m\bk+\bq}-\omega_{\bq\nu})^2}
+ \frac{n_{\bq\nu}+f_{m\bk+\bq}}
{(E_{\rm F}-\epsilon_{m\bk+\bq}+\omega_{\bq\nu})^2}
\right]\,,
\label{equation:lambda_abinitio}
\end{eqnarray}
and
\begin{eqnarray}
&&\Gamma_{\bq\nu} 
= 4\pi\sum_{m,n} \, \int \, \frac{d\bk}{A_{\rm BZ}}
\; |g_{mn,\nu}(\bk,\bq)|^2 \nonumber \\
&&\times\left( f_{n\bk} - f_{m\bk+\bq}\right)\ 
\delta\left(\epsilon_{m\bk+\bq}- \epsilon_{n\bk}-\omega_{\bq\nu}\right)\ .
\label{equation:gamma_abinitio}
\end{eqnarray}
Here $\epsilon_{n\bk}$ and $\omega_{\bq\nu}$ are the
energy eigenvalue of an electron with band index $n$ and
wavevector $\bk$ and that of a phonon with branch index $\nu$ and
wavevector $\bq$, respectively.
$A_{\rm BZ}$ is the area of the first Brillouin zone where
the integration is performed. The quantities $f_{n\bk}$ and $n_{\nu\bq}$ are the Fermi-Dirac
and the Bose-Einstein factors, respectively, and
$g_{mn,\nu}(\bk,\bq)\equiv\left<m\bk+\bq\right|\Delta V_{\nu\bq}(\br)\left|n\bk\right>$
is the scattering amplitude of an electronic state $\left|n\bk\right>$
into another state $\left|m\bk+\bq\right>$ resulting from the change
in the self-consistent field potential $\Delta V_{\nu\bq}(\br)$ arising from
a phonon with the branch index $\nu$ and the wavevector $\bq$.

\begin{figure}
\begin{center}
\includegraphics[width=1.0\columnwidth]{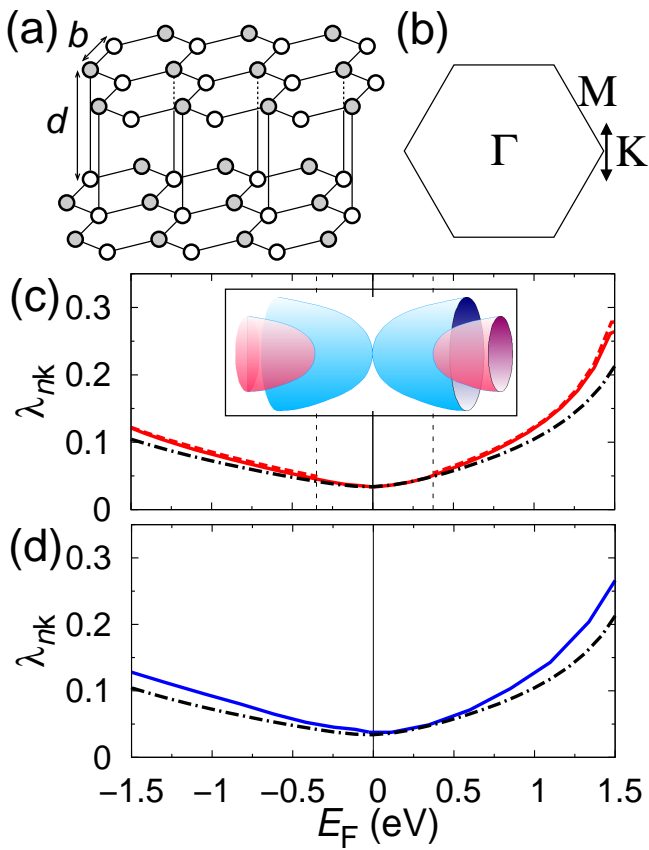}
\caption{(a) Ball-and-stick model of bilayer graphene (Bernal stacking).
(b) Brillouin zone of graphene and bilayer graphene.
(c) The electron-phonon
coupling strength $\lambda_{n\bk}$ in
bilayer graphene versus changing Fermi level $E_{\rm F}$
calculated along the path (double-head arrow) shown in (b).
Solid and dashed red lines correspond to $\lambda_{n\bk}$ of the individual
blue and red parabolic band in the inset, respectively.
The Fermi level of neutral bilayer graphene is set at zero.
(d) As in (c), for each of the two electronic
bands of {\it graphite} touching at the K point (solid blue line).
In (c) and (d), we show for comparison the \EP\ coupling strength in
graphene~\cite{park:2008PRB_Graphene_ElPh} (indicated by the dash-dotted line).}
\label{Figure_Lambda}
\end{center}
\end{figure}

Electron wavefunctions and energy eigenvalues are obtained
using {\it ab initio} pseudopotential density functional theory calculations~\cite{cohenlouie}
within the local density approximation~\cite{ceperley:1980PRL_pseudopotential,
perdew:1981PRB_exchcorr}.
Phonon frequencies and eigenstates are obtained
through density functional perturbation theory~\cite{baroni:2001RMP_DFPT}.
We have used a planewave basis set~\cite{ihm:1979JPC_PW}
with a kinetic energy cutoff
of 60~Ry. The core-valence interaction is handled using
norm-conserving pseudopotentials~\cite{troullier:1991PRB_pseudopotential,
fuchs:1999CPC_FHI98}.
The integration in Eq.~(\ref{equation:lambda_abinitio})
for graphene and bilayer graphene is performed by summation over
300$\times$300 points and for graphite it is performed with
90$\times$90$\times$30 points in the irreducible part of the Brillouin zone.
The integration of Eq.~(\ref{equation:gamma_abinitio}) for
graphene and bilayer graphene is done by summation over
1000$\times$1000 points in the irreducible part of the
Brillouin zone, and the $\delta$ function is replaced by a Lorentzian
with 15~meV broadening for convergence.
Electron and
phonon wavefunctions, energy eigenvalues and the \EP\ coupling matrix
elements in these extremely dense grid sets are obtained by
a recently developed interpolation scheme~\cite{giustino:2007PRL_Wannier,giustino:2007TBP_Migdal}
based on maximally localized Wannier
functions~\cite{marzari:1997PRB_Wannier,souza:2001PRB_Wannier}.
The Fermi-Dirac and Bose-Einstein factors
are evaluated at the temperature $T=15$~K in all the calculations.
Charge doping is modelled by adding or removing electrons from
the simulation cell and by using a neutralizing background.
In this work we assume that the layers in bilayer graphene and in graphite
are arranged according to the Bernal stacking sequence [Fig.~\ref{Figure_Lambda}(a)]~\cite{note:relax}.

Figure~\ref{Figure_Lambda} shows
the \EP\ coupling strengths $\lambda_{n\bk}$
in graphene, bilayer graphene and in graphite calculated
along the reciprocal space path indicated by the double-head arrow
in Fig.~\ref{Figure_Lambda}(b).
As pointed out in Ref.~\cite{park:2008PRB_Graphene_ElPh}
the \EP\ coupling strength
in graphene $\lambda_{n\bk}$ is insensitive to the location of the wavevector $\bk$
on the Fermi surface.
This is also the case for bilayer graphene and
for graphite. Therefore, we drop the index $n$ and the wavevector $\bk$ from now on.
In bilayer graphene, the two
electronic bands near the Dirac point energy
exhibit almost identical \EP\ coupling strengths [Fig.~\ref{Figure_Lambda}(c)].

The key factors determining
the \EP\ coupling strength are the density of states around the Dirac point energy
and the \EP\ matrix elements between the initial and the final electronic states
close to the Fermi level~\cite{grimvall:1981_Metal_ElPh}.
The density of states of pristine graphene vanishes at the Fermi level,
whereas bilayer graphene has a finite density of states.
Despite this difference,
at low doping with only one band occupied ($|E_{\rm F}-E_{\rm D}|<0.2$~eV, $E_{\rm D}$
being the energy at the Dirac point and is set to $E_{\rm D}=0$
in the following discussion),
the \EP\ coupling strengths in bilayer graphene and in graphite
are similar (within 5\%) to those of graphene.
This indicates that, as in graphene, there is no significant scattering
between low-energy electronic states
in bilayer graphene and in graphite arising from the \EP\ interaction.
This behavior originates from the chiral nature
of the charge carriers in bilayer
graphene~\cite{mccann:086805,katsnelson:2006NatPhys_Graphene_Klein}
and in graphite~\cite{wallace:1947PR_BandGraphite}, i.\,e.\,, it is a matrix-element effect.
The difference in the \EP\ coupling strength between graphene and bilayer graphene
(or graphite) increases with doping.
At the largest doping level considered
[$E_{\rm F}=1.5$~eV, Figs.~\ref{Figure_Lambda}(c) and
\ref{Figure_Lambda}(d)],
the coupling strength in bilayer graphene
and graphite ($\lambda=0.28$) is 30\%
larger than in graphene ($\lambda=0.21$).
As we show in the following, these differences result from
interlayer interaction.

\begin{figure*}
\includegraphics[width=2.0\columnwidth]{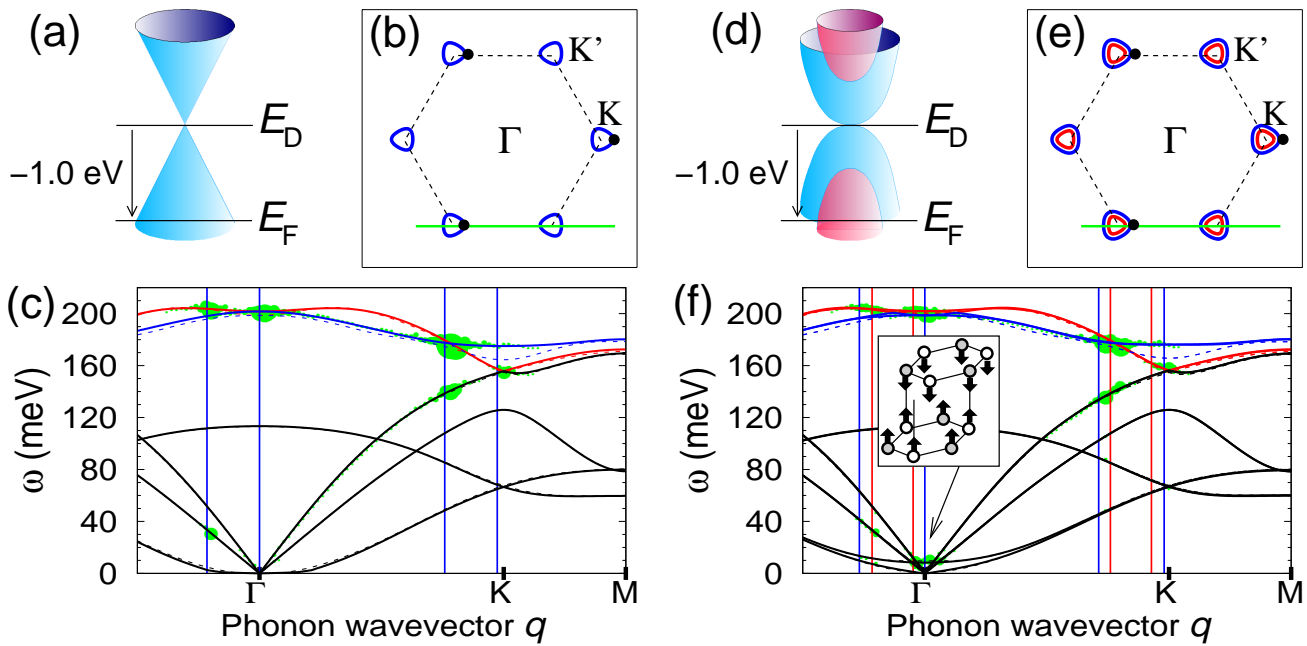}
\caption{
(a) The electronic energy dispersion and the Fermi level of hole-doped graphene.
(b) The Fermi surface (contours) and the Brillouin zone (dashed hexagon)
of hole-doped graphene.
The black dots represent the wavevector $\bk$ of the electronic state
considered on the Fermi surface.
(c) The phonon dispersion curves of undoped (dashed lines)
and hole-doped (solid lines) graphene
versus the wavevector $\bq$ along the solid green line shown in (b).
The vertical lines indicate the phonon wavevectors $\bq$ such that the
final electronic state with wavevector $\bk+\bq$ is also on the Fermi surface.
The size of the disks on top of the phonon dispersions is proportional to
the contribution of that phonon mode to $\lambda_{n\bk}$.
(d) to (f): Same quantities as in (a) to (c) for hole-doped bilayer graphene
but including also interband coupling.
The inset of (f) shows one of the three
modes responsible for the enhancement of
the \EP\ coupling strength in bilayer graphene.
The color (red and blue) and the type (solid and dashed) of the curves
in (c) and (f) corresponds to the phonon branches in Figs.~\ref{Figure_Gamma}(a) and
\ref{Figure_Gamma}(b)
and \ref{Figure_Gamma}(c) and \ref{Figure_Gamma}(d), respectively.}
\label{Figure_Lambda_decompose}
\end{figure*}

In order to determine which phonon modes lead to
the differences in the \EP\ coupling strengths
between monolayer and bilayer graphene,
we decomposed the coupling strength $\lambda_{n\bk}$ of both systems
into contributions from each phonon branch and wavevector.
Figures~\ref{Figure_Lambda_decompose}(b) and \ref{Figure_Lambda_decompose}(e)
show the Fermi surfaces of hole-doped  graphene and bilayer graphene
and the initial wavevector $\bk$ of the electronic state considered.
Figures~\ref{Figure_Lambda_decompose}(c) and \ref{Figure_Lambda_decompose}(f)
show the phonon dispersions of both pristine
and hole-doped graphene and bilayer graphene, respectively.
The size of the disks superimposed to the phonon dispersions is proportional to
the contribution to the coupling strength
$\lambda_{n\bk}$ arising from the corresponding
phonon mode. Figures~\ref{Figure_Lambda_decompose}(c) and
\ref{Figure_Lambda_decompose}(f)
show that both in graphene and in bilayer graphene,
the major contributions result from the highest-energy in-plane vibrations
with wavevectors connecting the initial and final electronic states
on the Fermi surface.
However, in the case of bilayer graphene, the three
low-energy optical branches with energy $\sim10$~meV enhance the \EP\ coupling strength
with respect to graphene.
The latter vibrations correspond to the compression mode (singly degenerate)
and to the sliding mode of the two layers
(doubly-degenerate).

Since, as in graphene, the \EP\ coupling strength in bilayer graphene is rather small,
even in the heavily doped case considered here,
it appears unlikely for bilayer graphene
to exhibit superconductivity with a transition temperature
significantly higher than that one may expect for graphene.

Recent angle-resolved photoemission experiments on
kish graphite~\cite{sugawara:036801} and on
a single crystal of graphite~\cite{leem:016802}
reported very different values of the \EP\ renormalization, namely,
$\lambda=0.70$ along the KK direction
[path indicated by the double-head arrow
in Fig.~\ref{Figure_Lambda}(b)]~\cite{sugawara:036801}
and $\lambda=0.14$ along the KM direction [Fig.~\ref{Figure_Lambda}(b)]~\cite{leem:016802}.
Our calculated \EP\ coupling strength in undoped graphite
($\lambda=0.034$)
is closer to the estimate of Ref.~\cite{leem:016802}.
In that work, the broadening of the energy distribution
curve in the photoemission spectra was entirely assigned to the
\EP\ interaction.
This assumption leads to an apparent \EP\
coupling strength which is enhanced by the contributions arising from other interactions.
In particular, it has been shown that the electronic linewidth in graphite arising
from the electron-electron interaction is sizable~\cite{PhysRevLett.87.246405}.
A similar discrepancy has been pointed out for the case of
graphene~\cite{gruneis:arXiv,mcchesney:arXiv2}. However, recent calculations
indicate that the effect of the electron-electron interaction in graphene is not
negligible and must be taken into account in the analysis of
the experimental data~\cite{park:elel_unp}.

\begin{figure}
\begin{center}
\includegraphics[width=1.0\columnwidth]{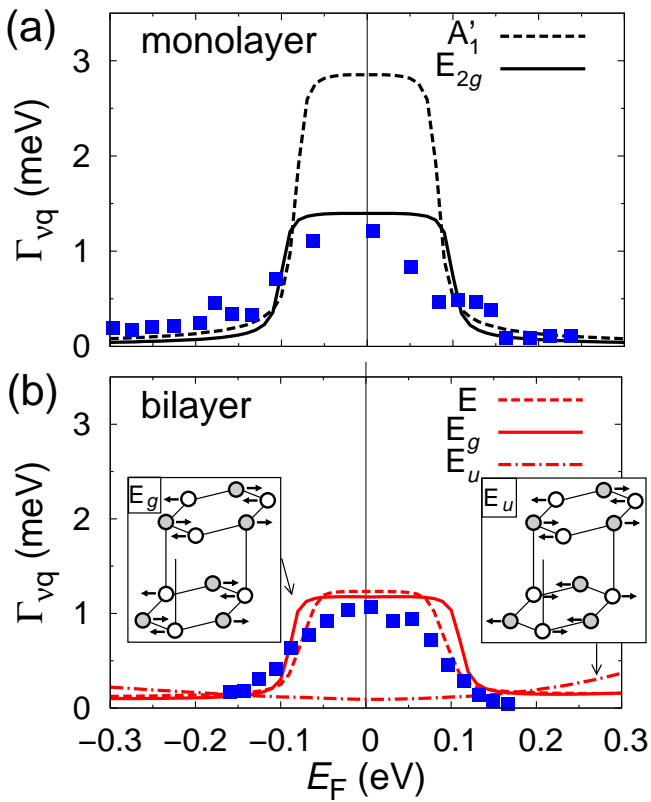}
\caption{(a) Phonon linewidth in doped graphene for
the ${\rm E}_{2g}$ mode at the $\Gamma$ point (solid line)
and the ${\rm A'}_1$ mode at the K point (dashed line)
versus the Fermi level $E_{\rm F}$.
The filled blue squares are the experimental data
from Ref.~\cite{yan:166802} downshifted
by 0.6~meV (to account for a uniform background).
(b) Phonon linewidth in bilayer graphene for the ${\rm E}_{g}$ mode
(solid line) and the ${\rm E}_{u}$ mode (dash-dotted line)
at the $\Gamma$ point, and for the E mode at the K point (dashed line).
The insets show one of each of
the two doubly-degenerate zone-center modes considered here.
The filled blue squares are the experimental data from
Ref.~\cite{yan:2007arxiv_bilayer} downshifted by 0.6~meV (to account for a uniform background).}
\label{Figure_Gamma_doping}
\end{center}
\end{figure}

So far we have discussed the effect of the \EP\ interaction
on the Fermi velocity of the carriers.
In what follows we focus on the effect of \EP\ interaction on
the phonon linewidths.
Figure~\ref{Figure_Gamma_doping}(a) shows the linewidth of the
doubly-degenerate ${\rm E}_{2g}$
phonons at the $\Gamma$ point
and of the doubly-degenerate ${\rm A'}_1$ mode at the K point for graphene.
These phonons exhibit a finite and constant linewidth
for $|E_{\rm F}|<\omega_{\rm ph}/2$, where
$\omega_{\rm ph}\sim0.2$~eV is the optical phonon energy,
and a negligible linewidth otherwise. The dependence of the phonon
linewidth on the doping level can be explained by
considering that interband transitions
through phonon absorption are forbidden whenever
$|E_{\rm F}|>\omega_{\rm ph}/2$.
Our calculated linewidth of the $E_{2g}$ phonon is in good agreement with
previous studies~\cite{ando:jpsj2006_graphene,lazzeri:266407}.

In bilayer graphene, as a consequence of the interlayer coupling,
the four highest-energy modes (originating from the ${\rm E}_{2g}$
modes of graphene) split into two sets of doubly-degenerate
${\rm E}_{g}$ and ${\rm E}_{u}$ modes~\cite{yan:125401},
with the ${\rm E}_{u}$ modes
1.1~meV higher in energy than the ${\rm E}_{g}$ modes
(cf. Fig.~\ref{Figure_Gamma_doping}).
Interestingly, at $\Gamma$, only the ${\rm E}_{g}$ modes exhibit a finite linewidth
(1.1~meV) whereas the ${\rm E}_{u}$ modes are not broadened
by the \EP\ interaction.
It can be shown that
this difference results from (i) the chiral nature of the low-energy electronic
states~\cite{wallace:1947PR_BandGraphite,Ando:JPSJ} and (ii) from the fact that atoms in the
same sublattice but different layers move in phase in the ${\rm E}_{g}$ modes, while
they move out of phase in the ${\rm E}_{u}$ modes~\cite{ando:2007_Bilayer}.
The calculated linewidths of the ${\rm E}_{g}$  and the ${\rm E}_{u}$ modes
are in good agreement with a previous study based on a pseudospin effective
Hamiltonian for the massive Dirac fermions of bilayer graphene~\cite{ando:2007_Bilayer}.

Among the two sets of high-energy zone-center modes,
only the ${\rm E}_{g}$ phonons are Raman active~\cite{yan:125401}.
The calculated phonon linewidth of the ${\rm E}_{g}$ modes can therefore
be compared directly with the measured broadening of the Raman lines.
As shown in Fig.~\ref{Figure_Gamma_doping}(b), our calculated linewidths are
in excellent agreement with those reported in
a recent experimental study~\cite{yan:2007arxiv_bilayer}.
For the purpose of comparison, we have downshifted the experimental
linewidths of the E$_{2g}$ phonon in graphene~\cite{yan:166802}
and of the E$_g$ phonon mode in bilayer graphene~\cite{yan:2007arxiv_bilayer} by 0.6~meV.
The agreement between our calculations and experiment after the subtraction of this uniform
background indicates that defect-induced scattering and
anharmonic effects
are small and similar in magnitude in graphene and bilayer graphene.

We note that the linewidth of the E$_{2g}$ phonons in graphene (1.1~meV)
and that of the E$_g$ phonons in bilayer graphene (1.4~meV) are very similar.
This behavior originates from the cancellation of the effects of larger electron density
of states and smaller \EP\ matrix elements of bilayer graphene as compared to graphene.
Because of their similar linewidths, the broadening of these modes
is unlikely to be useful for determining the number of graphene
layers using Raman spectroscopy.
In contrast, the linewidth of the highest-energy mode
at the K point in graphene (the A$_1'$ mode) is reduced from 2.9~meV in graphene
to 1.2~meV in bilayer graphene (the E mode).
Therefore it should be possible, at least in principle,
to exploit this difference in two-phonon Raman experiments to
distinguish between graphene and bilayer graphene.

\begin{figure*}
\includegraphics[width=2.0\columnwidth]{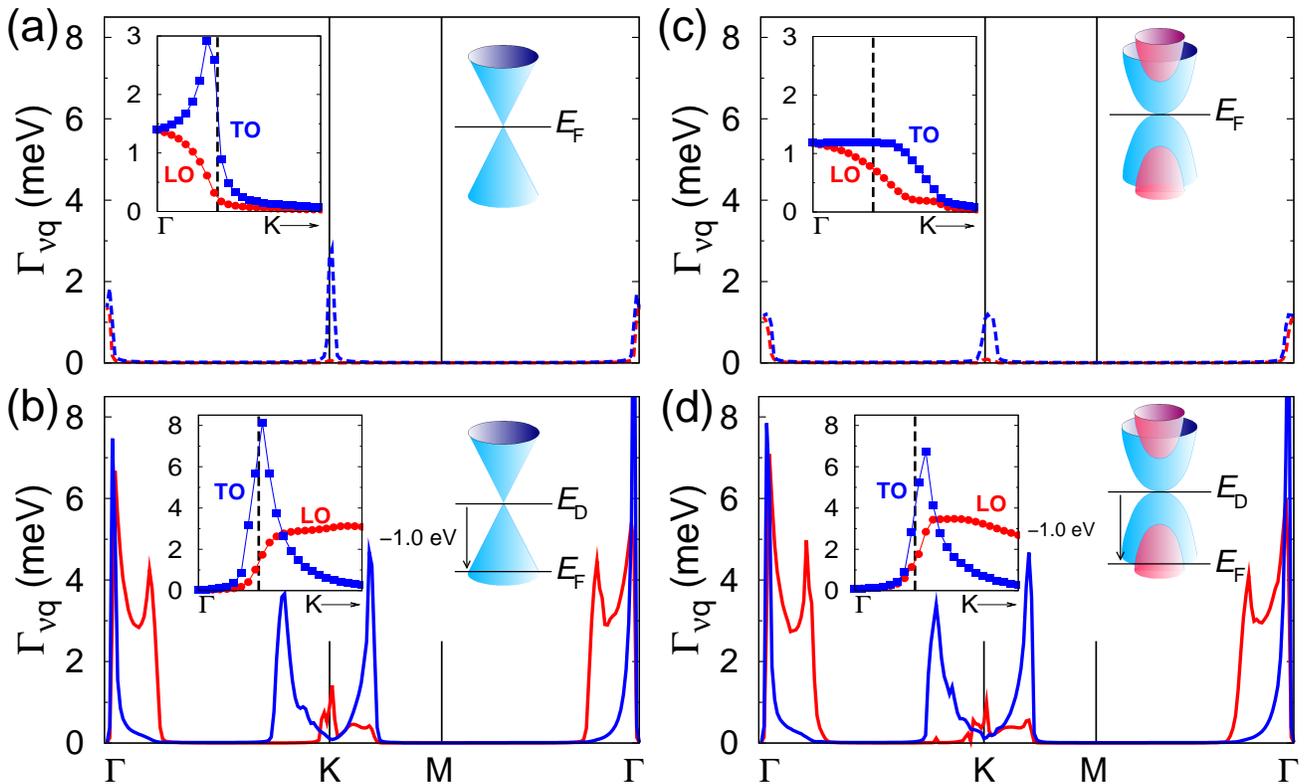}
\caption{(a), (b): The phonon linewidth of the highest energy branches
in undoped [(a)] and hole-doped
[(b)] graphene.
The color code (red and blue) and the type (solid and dashed) of the line
correspond to the phonon branches
shown in Fig.~\ref{Figure_Lambda_decompose}(c).
(c), (d): Phonon linewidth of the second highest doubly-degenerate
phonon branches at the $\Gamma$ point in undoped [(c)]
and hole-doped [(d)] bilayer graphene.
The color code (red and blue) and the type (solid and dashed) of the line
corresponds to the phonon branches
shown in Fig.~\ref{Figure_Lambda_decompose}(f).
The inset in each panel shows a magnified view of the region near the $\Gamma$ point,
where the symbols represent calculated data points and the lines are a guide to the eye.
The vertical dashed line in the inset specifies the characteristic wavevector
$k_{\rm 0}$ (see text).}
\label{Figure_Gamma}
\end{figure*}

Figure~\ref{Figure_Gamma}(a) shows
the linewidths of the two highest-energy phonon branches
in pristine graphene.
The phonon linewidths exhibit maxima at the
K point and at or near the $\Gamma$ point.
At the $\Gamma$ point,
the highest-energy phonons decay through electronic
transitions with no momentum transfer. Off the $\Gamma$ point, because of the
topology of the Dirac cone, non-vertical transitions can occur
if the wavevector of the phonon is smaller than
$k_0=\omega_{\rm ph}/v_{\rm F}=0.035~{\text \AA}^{-1}$
($\omega_{\rm ph}$ being the phonon energy and $v_{\rm F}$
the Fermi velocity).
These transitions are allowed since the phonon wavevectors connect
electronic states of the same chirality~\cite{piscanec:2004PRL_Graphite_KohnAnomaly}.
The scattering of phonons with wavevector
$k_0=\omega_{\rm ph}/v_{\rm F}$ is enhanced because
the phase velocity of the phonon matches the slope of the Dirac cone.
Correspondingly, at this wavevector the transverse-optical
phonon branch exhibits the largest linewidth.
Unlike the case for the transverse-optical phonons,
the longitudinal-optical phonons with wavevector $k_0$
cannot promote electronic transitions,
as a consequence of the chiral symmetry~\cite{piscanec:2004PRL_Graphite_KohnAnomaly},
and the corresponding linewidth vanishes.

As shown in Fig.~\ref{Figure_Gamma}(b),
in hole-doped graphene, the phonon linewidths
in the highest energy branches with wavevector
at the $\Gamma$ point or at the K point are negligible.
However, whenever the phonon wavevector exceeds $k_0$ in magnitude,
intraband electronic transitions can occur through
phonon absorption.
In the case of phonons with wavevector close to the K point,
this kind of electronic transition is suppressed
due to chirality~\cite{piscanec:2004PRL_Graphite_KohnAnomaly}.
In hole-doped graphene, the Fermi surface consists of two contours
centered around the two inequivalent Dirac points.
Two maxima are found in the phonon linewidths near K,
corresponding to the smallest and the largest
wavevectors connecting electronic states
on different contours.

Figures~\ref{Figure_Gamma}(c) and \ref{Figure_Gamma}(d) show
the linewidths of the E$_g$ phonons
for undoped as well as for hole-doped ($E_{\rm F}=-1.0$~eV)
bilayer graphene, respectively.
In the undoped case, the linewidths of transverse phonons
with wavevector $k_0$ are not as large as in graphene
since the low-energy electronic energy dispersions are nonlinear.
In the case of hole-doped bilayer graphene,
the profile of the phonon linewidths
is almost identical to the one calculated for graphene.
This can be explained by considering
that the electronic density of states per carbon atom near the Fermi level
in graphene and bilayer graphene are very similar.

The results of our calculations could be confirmed by performing
detailed inelastic neutron scattering,
electron energy loss spectroscopy,
or inelastic x-ray scattering
experiments.
Measurements of this kind have been performed,
for example, in the case of magnesium diboride, allowing for a direct comparison
with theoretical calculations.~\cite{PhysRevLett.90.095506}.

In summary the Fermi velocity renormalization
and the phonon line broadening arising from the \EP\ interaction
in bilayer graphene and in graphite are studied
and compared with the corresponding quantities
in graphene. In bilayer graphene and in graphite, the \EP\ coupling strength
is enhanced by up to 30\% at high doping as compared to graphene.
The calculated doping dependence of the phonon linewidth
of the zone-center E$_g$ mode in bilayer graphene is in excellent agreement
with recent Raman measurements~\cite{yan:166802}.
We discussed the similarities and the differences in the
linewidths of the optical phonons in graphene and in bilayer
graphene.

We thank E. Rotenberg, Jessica L. McChesney, Aaron Bostwick,
Michel C\^{o}t\'{e}, Jia-An Yan
and Jay Deep Sau  for fruitful discussions.
This work was supported by NSF Grant
No. DMR07-05941 and by the Director, Office of Science, Office of Basic Energy
Sciences, Division of Materials Sciences and Engineering Division,
U.S. Department of Energy under Contract No. DE- AC02-05CH11231.
Computational resources have been provided by
the National Partnership for Advanced Computational Infrastructure
(NPACI) and
the National Energy Research Scientific Computing Center (NERSC).
Part of the calculations were performed using modified versions of the
{\tt Quantum-Espresso}~\cite{baroni:2006_Espresso}
and the {\tt Wannier}~\cite{mostofi:2006_Wannier} packages.

{\it Additions and Corrections.}
In this paper [C.-H. Park, F. Giustino, M.L. Cohen, and S.G. Louie,
Nano Lett. {\bf 8}, 4229--4233 (2008)],
we failed to cite three relevant experimental
papers~\cite{pisana:nmat2007,das:nnano2008,das:arxiv2008}.
These references report on Raman measurements on
electrically gated single-layer graphene~\cite{pisana:nmat2007,das:nnano2008}
and bilayer graphene~\cite{das:arxiv2008}.

$^\dagger$Present address: University of Oxford,
Department of Materials, Parks Road, OX1 3PH, UK.

\end{document}